\documentclass[11pt,a4paper]{article}
\usepackage{jheppub}

\usepackage[latin1]{inputenc} 
\usepackage[T1]{fontenc}
\usepackage[safe]{textcomp}
\usepackage{lmodern} 
\usepackage{epsfig}
\usepackage{float}
\usepackage{amstext}  
\usepackage{amsfonts}
\usepackage{amssymb} 
\usepackage{amsmath}
\usepackage{amsthm}
\usepackage{bm}       
\usepackage[bottom]{footmisc} 
\usepackage{array}                
\usepackage{xcolor} 
\usepackage{framed}
\usepackage{slashed}
\usepackage[absolute]{textpos}
\usepackage{axodraw4j}
\usepackage{dsfont}
\usepackage{multirow}

\usepackage{appendix}
\usepackage{listings}
\usepackage{enumerate}     
\usepackage[bottom]{footmisc} 
\usepackage{array}           
\usepackage{parskip}        
\usepackage{xcolor} 
\usepackage{color}
\usepackage{framed}
\usepackage{fancyhdr}
\usepackage{subfig}

\usepackage{tikz}
\usepackage[absolute]{textpos}
\usepackage{pstricks}
\usepackage{axodraw4j}
\usepackage{graphicx}

\newcommand{\p}{\partial}

\newcommand{\f}[2]{\frac{#1}{#2}}
\newcommand{\sss}[1]{\scriptscriptstyle{#1}}
\newcommand{\ssst}[1]{\scriptscriptstyle{\text{#1}}}

\newcommand{\bea}{\begin{eqnarray}}
\newcommand{\eea}{\end{eqnarray}}
\newcommand{\be}{\begin{equation}}
\newcommand{\ee}{\end{equation}}
\newcommand{\ba}{\begin{align}}
\newcommand{\ea}{\end{align}}
\newcommand{\beas}{\begin{eqnarray*}}
\newcommand{\eeas}{\end{eqnarray*}}
\newcommand{\bes}{\begin{equation*}}
\newcommand{\ees}{\end{equation*}}
\newcommand{\bas}{\begin{align*}}
\newcommand{\eas}{\end{align*}}
 
 \newcommand{\Rig}{\Rightarrow}
\newcommand{\ssL}{{\mathcal L}} 
\newcommand{\eps}{{\varepsilon}}
 
\newcommand{\cf}{C_{\scriptscriptstyle{F}}} 
\newcommand{\ca}{C_{\scriptscriptstyle{A}}}
\newcommand{\tr}{T_{\scriptscriptstyle{F}}}

\newcommand{\Ng}{N_{\scriptscriptstyle{A}}}
\newcommand{\Ngl}{n_{\scriptscriptstyle{\tilde{g}}}}

\newcommand{\gs}{g_{\scriptscriptstyle{s}}}

\newcommand{\lb}{\left(}
\newcommand{\rb}{\right)}

\newcommand{\msbar}{$\overline{\text{MS}}$}

\newcommand{\dAAfNgex}{\frac{d_{\scriptscriptstyle{A}}^{abcd}d_{\scriptscriptstyle{A}}^{abcd}}{\Ng}}
\newcommand{\dFFfijNgex}{\frac{d_{\scriptscriptstyle{F,i}}^{abcd}d_{\scriptscriptstyle{F,j}}^{abcd}}{\Ng}}
\newcommand{\dFAfiNgex}{\frac{d_{\scriptscriptstyle{F,i}}^{abcd}d_{\scriptscriptstyle{A}}^{abcd}}{\Ng}}
\newcommand{\dFFfiNRex}{\frac{d_{\scriptscriptstyle{F,r}}^{abcd}d_{\scriptscriptstyle{F,i}}^{abcd}}{\dFr}}
\newcommand{\dFAfNRex}{\frac{d_{\scriptscriptstyle{F,r}}^{abcd}d_{\scriptscriptstyle{A}}^{abcd}}{\dFr}}

\newcommand{\dAAfNg}{d^{\sss{(4)}}_{\sss{AA}}}
\newcommand{\dFFfijNg}{d^{\sss{(4)}}_{\sss{FF,ij}}}
\newcommand{\dFAfiNg}{d^{\sss{(4)}}_{\sss{FA,i}}}
\newcommand{\dFFfiNR}{\tilde{d}^{\sss{(4)}}_{\sss{FF,ri}}}
\newcommand{\dFAfNR}{\tilde{d}^{\sss{(4)}}_{\sss{FA,r}}}

\newcommand{\cfi}{C_{\scriptscriptstyle{F,i}}} 
\newcommand{\cfj}{C_{\scriptscriptstyle{F,j}}} 
 
\newcommand{\cfr}{C_{\scriptscriptstyle{F,r}}} 

\newcommand{\tri}{T_{\scriptscriptstyle{F,i}}}

\newcommand{\trj}{T_{\scriptscriptstyle{F,j}}}

\newcommand{\trk}{T_{\scriptscriptstyle{F,k}}}

\newcommand{\trr}{T_{\scriptscriptstyle{F,r}}}
\newcommand{\dFr}{d_{\scriptscriptstyle{F,r}}}

\newcommand{\Nf}{n_{\scriptscriptstyle{f}}}
\newcommand{\Nfi}{n_{\scriptscriptstyle{f,i}}}
\newcommand{\Nfj}{n_{\scriptscriptstyle{f,j}}}
\newcommand{\Nfk}{n_{\scriptscriptstyle{f,k}}}
\newcommand{\Nfr}{n_{\scriptscriptstyle{f,r}}}

\definecolor{bluemar}{rgb}{0,0,.5}
\definecolor{redmar}{rgb}{.8,0,0}
\definecolor{greenmar}{rgb}{0,.5,0}

\newcommand{\bigrint}{%
\hbox to 2.92em{\hss\scalebox{1.1}[1] {\rotatebox[origin=c]{15}{$\displaystyle\int$}}\hss}}

\catcode`\@=11
\def\slash{\mathpalette\make@slash}
\def\make@slash#1#2{\setbox\z@\hbox{$#1#2$}%
  \hbox to 0pt{\hss$#1/$\hss\kern-\wd0}\box0}
\catcode`\@=12 

\allowdisplaybreaks
\newcommand{\ice}[1]{\relax}

\title{Four-loop renormalization of QCD with a reducible fermion representation
of the gauge group: anomalous dimensions and renormalization constants}
\author[a,b]{K.~G.~Chetyrkin} 
\author[c]{, M.~F.~Zoller}

\affiliation[a]{Institut f\"ur Theoretische Teilchenphysik, Karlsruhe Institute of Technology (KIT), Wolfgang-Gaede-Stra\ss{}e 1, 76131 Karlsruhe, Germany}

\affiliation[b]{II Institut f\"ur Theoretische Physik,  Universit\"at Hamburg,  Luruper Chaussee 149, 22761 Hamburg, Germany}

\affiliation[c]{Institut f\"ur Physik, Universit\"at Z\"urich (UZH), Winterthurerstrasse 190, 8057 Z\"urich, Switzerland}

\emailAdd{konstantin.chetyrkin@kit.edu}
\emailAdd{zoller@physik.uzh.ch}

\abstract{We present analytical results at four-loop level for the renormalization constants and anomalous dimensions of an extended QCD model
with one coupling constant and an arbitrary number of fermion representations. One example of such a model is the QCD plus gluinos sector of a
supersymmetric theory where the gluinos are Majorana fermions in the adjoint representation of the gauge group.

The renormalization constants of the gauge boson (gluon), ghost and fermion fields are analytically computed as well as those for the ghost-gluon 
vertex, the fermion-gluon vertex and the fermion mass. All other renormalization constants can be derived from these. Some of these results were 
already produced in Feynman gauge for the computation of the $\beta$-function of this model, which was recently published 
\cite{Zoller:2016sgq}.
Here we present results for an arbitrary $\xi$-parameter.
}

\keywords{Renormalization Group, QCD}
\arxivnumber{}
\subheader{TTP17-016, ZU-TH-07/17}

\begin{document}
\maketitle

\section{Introduction}

The behaviour of Green's functions wrt a shift of the renormalization scale is
described by the anomalous dimensions of the fields and parameters of the
theory, which enter the Renormalization Group Equations (RGE). For QCD the
full set of four-loop renormalization constants and anomalous dimensions was
presented in \cite{Chetyrkin:2004mf}.  The results for the four-loop QCD
$\beta$-function \cite{4loopbetaqcd,Czakon:2004bu} and the four-loop quark
mass and field anomalous dimensions had already been available
\cite{Vermaseren:1997fq,Chetyrkin:1997dh,Chetyrkin:1999pq}.
\footnote{Recently, the five-loop QCD $\beta$-function has been obtained  for
QCD colour factors \cite{Baikov:2016tgj} as well  as for a generic gauge group 
\cite{Herzog:2017ohr} (see, also,  \cite{Luthe:2016ima}).}

In this paper we consider a model with a non-abelian gauge group,
one coupling constant and a reducible fermion representation, i.~e.~any number of irreducible fermion representations.
The $\beta$-function for the coupling this model was computed in an earlier work \cite{Zoller:2016sgq}. Here we provide the remaining Renormalization Group (RG) functions
in full dependence on the gauge parameter $\xi$.

Apart from completing the set of renormalization constants and the RGE of the theory, which is important in itself, 
the gauge boson and ghost propagator anomalous dimensions serve another purpose. These quantities are essential ingredients in comparing the momentum
dependence of the corresponding propagators derived in non-perturbative calculations on the lattice, with perturbative results
(see 
e.~g.~\cite{Suman:1995zg,Becirevic:1999hj,Becirevic:1999uc,Becirevic:1999sc,vonSmekal:2009ae,Blossier:2010ph,Blossier:2011wk,Bornyakov:2013pha}).

This paper is structured as follows: First, we will give the notation and definitions for the model and the computed RG functions
We will also repeat how the special case of QCD plus Majorana gluinos in the adjoint representation of the gauge group can be derived from our
more general results.
Then we will present analytical results for the four-loop anomalous dimensions of the gauge boson, ghost and fermion field as well as the ones for the ghost-gluon 
vertex, the fermion-gluon vertex and the fermion mass in Feynman gauge for compactness. The renormalization constants and
anomalous dimensions for a generic gauge parameter $\xi$ can be found in machine readable form in an accompanying file, which can be downloaded
together with our source files on \texttt{www.arxiv.org}.

\section{Notation and definitions}

\subsection{QCD with several fermion representations}

The Lagrangian of a QCD-like model extended to include several fermion representations of the gauge group is given by
\bea
\ssL_{\sss{QCD}}&=&-\f{1}{4}G^a_{\mu \nu} G^{a\,\mu \nu}-\f{1}{2 \lambda}\lb\p_\mu A^{a\,\mu}\rb^2 
+\p_\mu \bar{c}^a \p^{\mu}c^a+\gs f^{abc}\,\p_\mu \bar{c}^a A^{b\,\mu} c^c \nonumber \\
&+&\sum\limits_{r=1}^{N_{\ssst{rep}}}\sum\limits_{q=1}^{\Nfr}
\left\{\f{i}{2}\bar{\psi}_{q,r}\overleftrightarrow{\slashed{\p}}\psi_{q,r}-m_{q,r}\bar{\psi}_{q,r}\psi_{q,r}
+ \gs \bar{\psi}_{q,r}\slashed{A}^a T^{a,r} \psi_{q,r}\right\}{},
\label{LQCD} 
\eea  
with the gluon field strength tensor
\be
G^a_{\mu \nu}=\p_\mu A^a_\nu - \p_\nu A^a_\mu + \gs f^{abc}A^b_\mu A^c_\nu{}.
\ee
The index $r$ specifies the fermion representation and the index $q$ the fermion flavour,
$\psi_{q,r}$ is the corresponding fermion field and $m_{q,r}$ the corresponding fermion mass. The number of fermion flavours in
representation $r$ is $\Nfr$ for any of the $N_{\ssst{rep}}$ fermion representations.

The generators $T^{a,r}$ of each fermion representation $r$ fulfill the defining anticomuting relation of the Lie Algebra corresponding to the gauge group:
\be \left[ T^{a,r},T^{b,r} \right]=if^{abc}T^{c,r}\ee
with the structure constants $f^{abc}$.
We have one quadratic Casimir operator $\cfr$ for each fermion representation, defined through
\be T^{a,r}_{ik} T^{a,r}_{kj} = \delta_{ij} \cfr, \ee
and $\ca$ for the adjoint representation. The dimensions of the fermion representations are given by $\dFr$ and the dimension of the adjoint representation
by $\Ng$. The traces of the different representations are defined as
\be \trr \delta^{ab}=\textbf{Tr}\lb T^{a,r} T^{b,r}\rb=T^{a,r}_{ij} T^{b,r}_{ji}. \ee
At four-loop level we also encounter higher order invariants in the gauge group factors which are expressed in terms of
symmetric tensors 
\be d_{\sss{R}}^{a_1 a_2 \ldots a_n}=\f{1}{n!} 
\sum\limits_{\text{perm } \pi}\text{Tr}\left\{ T^{a_{\pi(1)},R}T^{a_{\pi(2)},R}\ldots T^{a_{\pi(n)},R}\right\}{}, \label{dRa1an}\ee
where $R$ can be any fermion representation $r$, noted as $R=\{F,r\}$, or the adjoint representation, $R=A$, where
$T^{a,A}_{bc}=-i\,f^{abc}$.

An important special case of this model is the QCD plus gluinos sector of a
supersymmetric theory where the gluinos are Majorana fermions in the adjoint representation of the gauge group. 
Here we have $N_{\sss{rep}}=2$ and
\be 
\parbox{0.8\textwidth}{
\begin{tabular}{llllll}
$n_{\scriptscriptstyle{f,1}}$ & $=$ & $\Nf,\qquad$ & $n_{\scriptscriptstyle{f,2}}$ & $=$ & $\f{\Ngl}{2},$\\
$T_{\scriptscriptstyle{F,1}}$ & $=$ & $\tr,\qquad$ & $T_{\scriptscriptstyle{F,2}}$ & $=$ & $\ca,$\\
$C_{\scriptscriptstyle{F,1}}$ & $=$ & $\cf,\qquad$ & $C_{\scriptscriptstyle{F,2}}$ & $=$ & $\ca,$
\end{tabular}
}
\ee
the factor $\f{1}{2}$ in front of the number of gluinos $\Ngl$ being a result of the Majorana nature of 
the gluinos (see e.~g.~\cite{Clavelli:1996pz}). This can be understood in the following way:
It has been shown in \cite{Denner:1992vza} that one can treat Majorana
fermions by first drawing all possible Feynman diagrams and choosing an arbitrary orientation (fermion flow) for each
fermion line. Then Feynman rules are applied in the same way as for Dirac spinors, especially one can use the same propagators
$\f{i}{\slashed{p}-m}$ for the momentum $p$ along the fermion flow and $\f{i}{-\slashed{p}-m}$ for $p$ against the fermion flow.
Closed fermion loops receive a factor $(-1)$. One then applies the same symmetry factors as for scalar or vector particles,
e.~g.~a factor $\f{1}{2}$ for a loop consisting of two propagators of Majorana particles. For this work
we generate our diagrams using one Dirac field $\psi$ for all fermions, i.~e.~we produce both possible fermion flows
in loops unless they lead to the same diagram. The latter case is exactly the one where the symmetry factor $\f{1}{2}$ must be applied.
The first case means that the loop was double-counted which should also be compensated by a factor $\f{1}{2}$.

By adding counterterms to the Lagrangian \eqref{LQCD} in order to remove all possible UV divergences we arrive at the bare Lagrangian expressed
through renormalized fields, masses and the coupling constant:
\bea
\ssL_{\sss{QCD,B}}&=&
-\f{1}{4} Z_3^{(2g)} \lb \p_\mu A^a_\nu - \p_\nu A^a_\mu \rb^2-\f{1}{2 \lambda}\lb\p_\mu A^{a\,\mu}\rb^2 \nonumber\\
&-&\f{1}{2} Z^{(3g)}_1 \gs f^{abc}\lb \p_\mu A^a_\nu 
- \p_\nu A^a_\mu \rb A^b_\mu A^c_\nu \nonumber\\
&-&\f{1}{4} Z^{(4g)}_1 \gs^2 \lb f^{abc} A^b_\mu A^c_\nu \rb^2
+ Z^{(2c)}_3 \p_\mu \bar{c}^a \p^{\mu}c^a
+ Z^{(ccg)}_1 \gs f^{abc}\,\p_\mu \bar{c}^a A^{b\,\mu} c^c \\
&+&\sum\limits_{r=1}^{N_{\ssst{rep}}}\sum\limits_{q=1}^{\Nfr}
\left\{Z^{(q,r)}_2\f{i}{2}\bar{\psi}_{q,r}\overleftrightarrow{\slashed{\p}}\psi_{q,r}
-m_{q,r} Z_{m}^{(q,r)} Z^{(q,r)}_2\bar{\psi}_{q,r}\psi_{q,r} \right.\nonumber \\
&+&\left. \gs Z^{(q,r)}_1\bar{\psi}_{q,r}\slashed{A}^a T^{a,r} \psi_{q,r}\right\}{},\nonumber
\label{LQCD0}
\eea
were we have already used the fact that $Z_\lambda=Z_3^{(2g)}$.

Due to the Slavnov-Taylor identities all vertex renormalization constants are connected and can be expressed through 
the renormalization constant of the coupling constant and the renormalization constants of the fields appearing in the respective vertex:
\bea 
Z_{\gs} &=& Z^{(3g)}_{1} \lb Z^{(2g)}_{3} \rb^{-\f{3}{2}}{}, \label{Zgscomputation_3g}\\
Z_{\gs} &=& \sqrt{Z^{(4g)}_{1}} \lb Z^{(2g)}_{3} \rb^{-1}{},\label{Zgscomputation_4g}\\
Z_{\gs} &=& Z^{(ccg)}_{1} \lb Z^{(2c)}_{3}\sqrt{Z^{(2g)}_{3}} \rb^{-1}{}, \label{Zgscomputation_ccg}\\
Z_{\gs} &=& Z^{(q,r)}_1 \lb Z^{(q,r)}_2\sqrt{Z^{(2g)}_{3}} \rb^{-1}{}. \label{Zgscomputation_qr}
\eea

In the \msbar-scheme using regularization in $D=4-2 \eps$ space time dimensions all renormalization constants have the form
\be Z(a,\lambda)=1+\sum\limits_{n=1}^{\infty} \f{z^{(n)}(a,\lambda)}{\eps^n}{}, \label{Zdef}\ee
where $a=\f{\gs^2}{16\pi^2}$. From the fact that the bare parameter $a_{\sss{B}}=Z_a a \mu^{2\eps}$ (with $Z_a=Z_{\gs}^2$)  
does not depend on the renormalization scale $\mu$ one finds
\bea \beta^{(D)}(a)&=& \mu^2\f{da}{d\mu^2}=-\eps a+\beta(a){},\\
\beta(a) &=& a^2 \f{d}{da} z^{(1)}_a(a){}.
\eea
Given a renormalization constant $Z$ the corresponding anomalous dimension is defined as
\be
\gamma(a,\lambda)=-\mu^2\f{d \log Z(a,\lambda)}{d\mu^2}=a \f{\partial z^{(1)}(a)}{\partial a}:=
-\sum\limits_{n=1}^{\infty} \gamma^{(n)}(\lambda)\,a^{n}{}.
\label{gammadef}
\ee
From the definition of anomalous dimensions \eqref{gammadef} it follows that
\be \gamma(a,\lambda)= \lb \eps a - \beta(a) \rb \f{d \log Z(a,\lambda)}{da}
- \gamma_3^{(2g)}(a,\lambda) \lambda \f{d \log Z(a,\lambda)}{d\lambda}{}, \label{gammaZcorr} \ee
where we use the fact that the evolution of any parameter (or field) -- here $\lambda$ -- is described by its anomalous dimension, i.~e.
\be \lambda_{\sss{B}}= Z_\lambda \lambda \; \Rig \; \mu^2\f{d}{d\mu^2}\lambda=\gamma_\lambda \lambda{},\ee
and the fact that $\gamma_\lambda=\gamma_3^{(2g)}$. Using \eqref{gammaZcorr} one can reconstruct renormalization constants from the
corresponding anomalous dimension, a finite and usually more compact quantity, and the $\beta$-function of the model.

\subsection{Technicalities}
The 1-particle-irreducible Feynman diagrams needed for this project
were generated with QGRAF \cite{QGRAF}.  We compute $Z^{(2c)}_3$,
$Z^{(2g)}_3$ and $Z^{(q,r)}_2$ from the 1PI self-energies of the
fields $A^a_\mu$, $c$ and $\psi_{q,r}$ as well as $Z^{(ccg)}_1$ and
$Z^{(q,r)}_1$ from the respective vertex corrections and
$Z_{m}^{(q,r)}$ from the 1PI corrections to a Green's function with an
insertion of one operator $\bar{\psi}_{q,r}\psi_{q,r}$ and an external
fermion line of type $(q,r)$.
We used two different methods to calculate these objects, first a direct four-loop calculation in Feynman gauge with massive tadpoles and then
an indirect method where four-loop objects are constructed from propagator-like three-loop objects to derive the full dependence on the gauge parameter $\xi:=1-\lambda$.

\subsubsection{Direct four-loop calculation in the Feynman gauge with massive tadpoles}

For $\xi=0$ (Feynman gauge) the topologies of the diagrams were
identified with the C++ programs Q2E and EXP
\cite{Seidensticker:1999bb,Harlander:1997zb}. In this approach all
diagrams were expanded in the external momenta in order to factor out
the momentum dependence of the tree-level vertex or propagator, e.~g.~$q^{\mu}q^{\nu}-q^2 g^{\mu\nu}$
for the gluon self-energy. Then the
tensor integrals were projected onto scalar integrals, using e.~g.~$\f{q^\mu q^\nu}{q^4}$
as well as $\f{g^{\mu\nu}}{q^2}$ as projectors
for the gluon self-energy. After this we set all external momenta to
zero since the UV divergent part of the integral does not depend on
finite external momenta. We then use the method of introducing the
same auxiliary mass parameter $M^2$ in every propagator denominator
\cite{Misiak:1994zw,beta_den_comp}. Subdivergencies $\propto M^2$ are
cancelled by an unphysical gluon mass counterterm
\mbox{$\f{M^2}{2}\delta\!Z_{\sss{M^2}}^{(2g)}\,A_\mu^a A^{a\,\mu}$}
restoring the correct UV divergent part of the diagrams.  This method
was e.~g.~used in
\cite{4loopbetaqcd,Schroder:2002re,DiRenzo:2004ws,Czakon:2004bu,Chetyrkin:2012rz,Zoller:2015tha,Chetyrkin:2016ruf}
and is explained in detail in \cite{Zoller:2014xoa}.

For the expansions, application of projectors, evaluation of fermion
traces and counterterm insertions in lower loop diagrams we used FORM
\cite{Vermaseren:2000nd,Tentyukov:2007mu}. The scalar tadpole
integrals were computed with the \mbox{FORM}-based package
\mbox{MATAD} \cite{MATAD} up to three-loop order. At four loops we use
the C++ version of FIRE 5 \cite{Smirnov:2008iw,Smirnov:2014hma} in
order to reduce the scalar integrals to Master Integrals which can be
found in \cite{Czakon:2004bu}. Technical details of the reduction are
described in the previous paper \cite{Zoller:2015tha}.

\subsubsection{Indirect four-loop calculation using three-loop massless propagators}

The case of a generic gauge parameter $\xi$ is certainly possible to treat in
the same \textit{massive} way but calculations then require significantly more time
and computer resources\footnote{Nevertheless, it has been done recently along
  theses lines in \cite{Luthe:2017ttc} for the case of one irreducible fermion
  representation.}.  As a result we have chosen an alternative \textit{massless}
approach which reduces the evaluation of any $L$-loop Z-factor to the calculation
of some properly constructed set of $(L-1)$-loop massless propagators
\cite{Vladimirov:1979zm,Chetyrkin:1984xa,Chetyrkin:2017ppe,Baikov:2015tea}.
 As is well-known (starting already from $L=2$ \cite{Davydychev:1992mt})
calculation of $L$-loop massive vacuum diagrams is significantly more complicated
and time-consuming than the one of corresponding $(L-1)$-loop massless
propagators.

The approach is easily applicable for any Z-factor except for $Z_3$
\cite{Chetyrkin:2004mf}. The latter problem is certainly doable within the
massless approach but requires significantly more human efforts in resolving
rather sophisticated combinatorics\footnote{Very recently the problem has been
successfully solved in two radically different ways \cite{Baikov:2016tgj} and
\cite{Herzog:2017ohr}.}. On the other hand, one could restore the full
$\xi$-dependence of $Z_3$ from all other renormalization constants and from the fact that the
charge renormalization constant $Z_g$ is gauge invariant 
\cite{Chetyrkin:2004mf,Luthe:2017ttc}. As $Z_g$ in QCD with fermions transforming under 
arbitrary reducible representation of the  gauge group has been recently found in  
\cite{Zoller:2016sgq} we  have  proceeded in this way. For calculation of 3-loop massless 
propagator we have used the FORM version of MINCER \cite{Larin:1991fz}.

\subsubsection{computation of the gauge group factors}

 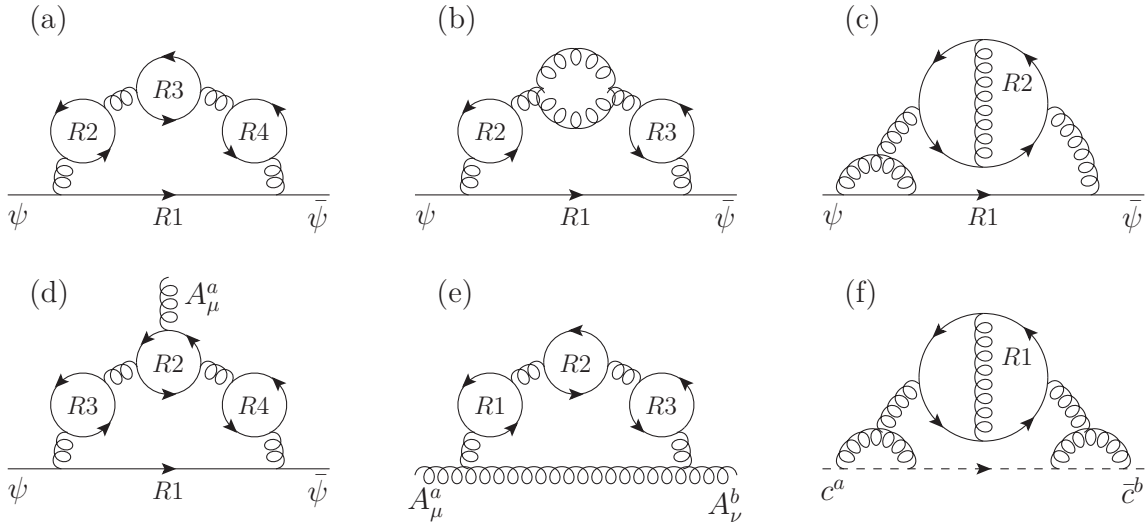
\begin{figure}[ht!]
 \begin{center}
  \begin{tabular}{ccc}
   \scalebox{0.8}{  \begin{picture}(140,100) (0,0)
    \SetWidth{0.5}
    \SetColor{Black}
    \ArrowLine(-5,0)(145,0)
    \GlueArc(70,0)(50,0,20){4}{2}
    \GlueArc(70,0)(50,54,73){4}{2}
    \GlueArc(70,0)(50,107,126){4}{2}
    \GlueArc(70,0)(50,160,180){4}{2}
     \ArrowArc(70,50)(15,0,180)
     \ArrowArc(70,50)(15,180,360)
     \ArrowArc(30,30)(15,45,225)
     \ArrowArc(30,30)(15,225,45)
     \ArrowArc(110,30)(15,-45,135)
     \ArrowArc(110,30)(15,135,315)
    \Text(0,-10)[cc]{\Large{\Black{$\psi$}}}
    \Text(140,-10)[cc]{\Large{\Black{$\bar{\psi}$}}}
    \Text(70,-10)[cc]{\large{\Black{$R1$}}}
    \Text(30,30)[cc]{\large{\Black{$R2$}}}
    \Text(70,50)[cc]{\large{\Black{$R3$}}}    
    \Text(110,30)[cc]{\large{\Black{$R4$}}}
    \Text(5,75)[lb]{\Large{\Black{(a)}}}
  \end{picture}} \qquad\quad
  &
     \scalebox{0.8}{  \begin{picture}(140,100) (0,0)
    \SetWidth{0.5}
    \SetColor{Black}
    \ArrowLine(-5,0)(145,0)
    \GlueArc(70,0)(50,0,20){4}{2}
    \GlueArc(70,0)(50,54,73){4}{2}
    \GlueArc(70,0)(50,107,126){4}{2}
    \GlueArc(70,0)(50,160,180){4}{2}
     \GlueArc(70,50)(15,0,180){4}{5}
     \GlueArc(70,50)(15,180,360){4}{5}
     \ArrowArc(30,30)(15,45,225)
     \ArrowArc(30,30)(15,225,45)
     \ArrowArc(110,30)(15,-45,135)
     \ArrowArc(110,30)(15,135,315)
    \Text(0,-10)[cc]{\Large{\Black{$\psi$}}}
    \Text(140,-10)[cc]{\Large{\Black{$\bar{\psi}$}}}
    \Text(70,-10)[cc]{\large{\Black{$R1$}}}
    \Text(30,30)[cc]{\large{\Black{$R2$}}}   
    \Text(110,30)[cc]{\large{\Black{$R3$}}}
    \Text(5,75)[lb]{\Large{\Black{(b)}}}
  \end{picture}} \qquad\quad
  &
   \scalebox{0.8}{  \begin{picture}(140,100) (0,0)
    \SetWidth{0.5}
    \SetColor{Black}
    \ArrowLine(-5,0)(145,0)
    \GlueArc(70,0)(50,0,53){4}{6}
    \GlueArc(70,0)(50,127,160){4}{4}
    \GlueArc(20,0)(15,0,180){4}{7}
     \ArrowArc(70,43)(30,0,90)
     \ArrowArc(70,43)(30,90,180)
     \ArrowArc(70,43)(30,180,270)
     \ArrowArc(70,43)(30,270,360)
     \Gluon(70,13)(70,73){4}{8}
    \Text(0,-10)[cc]{\Large{\Black{$\psi$}}}
    \Text(140,-10)[cc]{\Large{\Black{$\bar{\psi}$}}}
    \Text(70,-10)[cc]{\large{\Black{$R1$}}}
    \Text(86,53)[cc]{\large{\Black{$R2$}}}    
    \Text(5,75)[lb]{\Large{\Black{(c)}}}
  \end{picture}}\\[4ex]
   \scalebox{0.8}{  \begin{picture}(140,100) (0,0)
    \SetWidth{0.5}
    \SetColor{Black}
    \ArrowLine(-5,0)(145,0)
    \Gluon(70,65)(70,90){4}{3}
    \GlueArc(70,0)(50,0,20){4}{2}
    \GlueArc(70,0)(50,54,73){4}{2}
    \GlueArc(70,0)(50,107,126){4}{2}
    \GlueArc(70,0)(50,160,180){4}{2}
     \ArrowArc(70,50)(15,0,90)
     \ArrowArc(70,50)(15,90,180)
     \ArrowArc(70,50)(15,180,360)
     \ArrowArc(30,30)(15,45,225)
     \ArrowArc(30,30)(15,225,45)
     \ArrowArc(110,30)(15,-45,135)
     \ArrowArc(110,30)(15,135,315)
    \Text(0,-10)[cc]{\Large{\Black{$\psi$}}}
    \Text(140,-10)[cc]{\Large{\Black{$\bar{\psi}$}}}
    \Text(86,80)[cc]{\Large{\Black{$A^{a}_{\mu}$}}}
    \Text(70,-10)[cc]{\large{\Black{$R1$}}}
    \Text(30,30)[cc]{\large{\Black{$R3$}}}
    \Text(70,50)[cc]{\large{\Black{$R2$}}}    
    \Text(110,30)[cc]{\large{\Black{$R4$}}}
    \Text(5,75)[lb]{\Large{\Black{(d)}}}
  \end{picture}} \qquad\quad
  &
   \scalebox{0.8}{  \begin{picture}(140,100) (0,0)
    \SetWidth{0.5}
    \SetColor{Black}
    \Gluon(-5,-3)(145,-3){4}{22}
    \GlueArc(70,0)(50,0,20){4}{2}
    \GlueArc(70,0)(50,54,73){4}{2}
    \GlueArc(70,0)(50,107,126){4}{2}
    \GlueArc(70,0)(50,160,180){4}{2}
     \ArrowArc(70,50)(15,0,180)
     \ArrowArc(70,50)(15,180,360)
     \ArrowArc(30,30)(15,45,225)
     \ArrowArc(30,30)(15,225,45)
     \ArrowArc(110,30)(15,-45,135)
     \ArrowArc(110,30)(15,135,315)
    \Text(0,-17)[cc]{\Large{\Black{$A^{a}_{\mu}$}}}
    \Text(140,-17)[cc]{\Large{\Black{$A^{b}_{\nu}$}}}
    \Text(30,30)[cc]{\large{\Black{$R1$}}}
    \Text(70,50)[cc]{\large{\Black{$R2$}}}    
    \Text(110,30)[cc]{\large{\Black{$R3$}}}
    \Text(5,75)[lb]{\Large{\Black{(e)}}}
  \end{picture}} \qquad\quad
  &
   \scalebox{0.8}{  \begin{picture}(140,100) (0,0)
    \SetWidth{0.5}
    \SetColor{Black}
    \DashArrowLine(-5,0)(145,0){4}
    \GlueArc(70,0)(50,20,53){4}{4}
    \GlueArc(70,0)(50,127,160){4}{4}
    \GlueArc(20,0)(15,0,180){4}{7}
    \GlueArc(120,0)(15,0,180){4}{7}
     \ArrowArc(70,43)(30,0,90)
     \ArrowArc(70,43)(30,90,180)
     \ArrowArc(70,43)(30,180,270)
     \ArrowArc(70,43)(30,270,360)
     \Gluon(70,13)(70,73){4}{8}
    \Text(0,-10)[cc]{\Large{\Black{$c^a$}}}
    \Text(140,-10)[cc]{\Large{\Black{$\bar{c}^b$}}}
    \Text(86,53)[cc]{\large{\Black{$R1$}}}    
    \Text(5,75)[lb]{\Large{\Black{(f)}}}
  \end{picture}}\\[4ex]
\end{tabular}\end{center}
\caption{Four-loop diagrams contributing to the fermion self-energy (a,b,c), the fermion-gauge-boson-vertex (d), the gluon self-energy (e)
and the ghost self-energy (f). Each fermion line is initially treated as a different representation $R1,\ldots,R4$.}
\label{dias_fermionlines} 
\end{figure}
The calculation of the gauge group factors was done with an extended
version of the \mbox{FORM} package COLOR \cite{COLOR} already used and
presented in \cite{Zoller:2016sgq}. We take the following steps:
\begin{enumerate}
 \item For the generation of the diagrams in QGRAF \cite{QGRAF} we use one field $A$ for the adjoint representation (gauge boson) and one field $\psi$ for all 
 the fermion representations. This has the advantage that we do not produce more Feynman diagrams than in QCD. Each fermion line in a diagram 
 gets a line number and is treated as a different representation from the other fermion lines. Since we compute diagrams up to four-loop order we
 need up to four different line representations $R1,\ldots,R4$ (see Fig.~\ref{dias_fermionlines}) with the
 generators $T^{a,R1}_{ij}=\texttt{T1(i,j,a)}$, 
$T^{a,R2}_{ij}=\texttt{T2(i,j,a)}$, $T^{a,R3}_{ij}=\texttt{T3(i,j,a)}$ and $T^{a,R4}_{ij}=\texttt{T4(i,j,a)}$. 
Each fermion loop gets assigned a factor $\Nf$.
\item The modified version of COLOR \cite{COLOR,Zoller:2016sgq} then writes the generators into traces 
\be \textbf{Tr} \left\{ T^{a_1,R} \ldots T^{a_n,R}\right\} = \texttt{TR\{R\}(a1,\ldots,an)}, \quad (R=R1,\ldots,R4) \ee
which are then reduced as outlined in \cite{COLOR} yielding colour factors expressed through traces \texttt{TF\{R\}},
the Casimir operators \texttt{cF\{R\}} and \texttt{cA}, the dimensions of the representations \texttt{dF\{R\}} and \texttt{NA}.
\item Now we change from fermion line numbers $R1,\ldots,R4$ to four explicit physical fermion representations $r$ by substituting each of the 
line numbers $R1,\ldots,R4$ by the sum over all representations $r=1,\ldots,4$. 
An example of the substitution of $\{R1,\ldots,R4\}$-colour factors with those of the physical representaions in a one-loop diagram is
\be \texttt{Nf*TF1} \to n_{\sss{f,1}} T_{\sss{F,1}} + n_{\sss{f,2}} T_{\sss{F,2}} + n_{\sss{f,3}} T_{\sss{F,3}}+ n_{\sss{f,4}} T_{\sss{F,4}}. \ee
At higher orders this subtitution becomes much more involved\footnote{For this reason it is convenient to collect all combinations $\texttt{Nf}^\texttt{x1}\texttt{*TF1}^\texttt{x2}\texttt{*CF1}^\texttt{x3}
\texttt{*TF2}^\texttt{x4}\texttt{*CF2}^\texttt{x5}\texttt{*TF3}^\texttt{x6}\texttt{*CF3}^\texttt{x7}\texttt{*TF4}^\texttt{x8}\texttt{*CF4}^\texttt{x9}$
in a function \texttt{C(x1,\ldots,x9)}. The factors \texttt{C(x1,\ldots,x7)} are then substituted by the proper 
combinations of $n_{\sss{f,1}}$, $T_{\sss{F,1}}$, $c_{\sss{F,1}}$, etc.}. 
Diagram (a) from Fig.~\ref{dias_fermionlines} now corresponds to a sum of $4^4=256$ diagrams with explicit fermion representations. 
This lengthy representation of our results is needed for the renormalization procedure, since e.~g.~a one loop counterterm to the gluon 
self-energy, computed from a diagram with only $R1$, must be applied to all the fermion loops in Fig.~\ref{dias_fermionlines} (a,b,d,e). 
This is most conveniently achieved if each fermion-loop is considered a sum over
all physical fermion representations just as it is considered a some over all (massless) fermion 
flavours.\footnote{Since renormalization constants in the \msbar-scheme do not depend on masses all fermion 
flavours can be treated as massless for their computation.} The factors involving $d_{\sss{F,r}}^{a_1 a_2 a_3 a_4}$, 
$d_{\sss{F,r}}^{a_1 a_2 a_3}$, $d_{\sss{A}}^{a_1 a_2 a_3 a_4}$ and $d_{\sss{A}}^{a_1 a_2 a_3}$ appear only at four-loop level and do hence
not interfere with lower order diagrams with counterterm insertions. They can be treated directly in the next step.
\item 
After all subdivergencies are cancelled by adding the lower-loop diagrams with counterterm insertions we simplify and generalize the notation.
The explicit colour factors are collected in sums of terms built from $\Nfr$, $\cfr$ and $\trr$ over all physical representations $r$, e.~g.\footnote{For convenience we collect
$n_{\sss{f,1}}^{x_1} n_{\sss{f,2}}^{x_2} n_{\sss{f,3}}^{x_3}  n_{\sss{f,4}}^{x_4}
 T_{\sss{F,1}}^{y_1} T_{\sss{F,2}}^{y_2} T_{\sss{F,3}}^{y_3}  T_{\sss{F,4}}^{y_4} 
 C_{\sss{F,1}}^{z_1} C_{\sss{F,2}}^{z_2} C_{\sss{F,3}}^{z_3} C_{\sss{F,4}}^{z_4}$ in a function
\texttt{CR(x1,\ldots,x4,y1,\ldots,y4,z1\ldots,44)} which are then substituted by the proper sums of colour factors over all representations $r$.}
\be n_{\sss{f,1}} T_{\sss{F,1}} \to \sum\Nfi\tri - n_{\sss{f,2}} T_{\sss{F,2}} - n_{\sss{f,3}} T_{\sss{F,3}}- n_{\sss{f,4}} T_{\sss{F,4}}. \ee
Since we used the maximum number of different fermion representations which can appear in any diagram the result is valid for any number of fermion
representations $N_{\sss{rep}}$. 
\end{enumerate}

\section{Results \label{res:beta}}
In this section we give the results for the anomalous dimensions of the QCD-like model with an arbitrary number of fermion representations as described above
to four-loop level. The number of active fermion flavours of representation $i$ is denoted by $\Nfi$. Apart from the Casimir operators $\ca$ and $\cfi$
and the trace $\tri$ the following invariants appear in our results:
\bea  \dAAfNg &=& \dAAfNgex,\quad \dFAfiNg= \dFAfiNgex,\quad  \dFFfijNg=\dFFfijNgex, \nonumber\\ \dFAfNR&=& \dFAfNRex,\quad  \dFFfiNR=\dFFfiNRex, \eea
where $r$ is fixed and $i,j$ will be summed over all fermion representations.
In this section we give the results for $\lambda=1$ (Feynman gauge),
the general case $\lambda=(1-\xi)$ can be found in the accompanying source files on \texttt{www.arxiv.org}.

From the gauge boson field strength renormalization constant $Z_{3}^{(2g)}$ we compute the anomalous dimension according to \eqref{gammadef}
\bea
\lb\gamma_{3}^{(2g)}\rb^{(1)}&=&  
       - \f{5}{3} \ca  + \sum\limits_i \f{4}{3}\Nfi \tri \label{1lgamma3g} {},       \\
\lb\gamma_{3}^{(2g)}\rb^{(2)}&=& 
      - \f{23}{4} \ca^2 + \sum\limits_i \Nfi\tri \lb 4 \cfi        + 5 \ca \rb          \label{2lgamma3g} {},       \\
\lb\gamma_{3}^{(2g)}\rb^{(3)}&=&     
       - \ca^3 \lb \f{4051}{144}
          - \f{3}{2} \zeta_{3} \rb
          + \sum\limits_i \Nfi \tri   \left[
          - 2 \cfi^2
          + \ca \cfi \lb \f{5}{18} + 24 \zeta_{3} \rb             \right. \nonumber \\  & &\left.
          + \ca^2 \lb \f{875}{18} - 18 \zeta_{3}  \rb        
          \right]
       - \sum\limits_{i,j}\Nfi \Nfj \tri \trj   \left(
           \f{44}{9} \cfj
          + \f{76}{9} \ca
          \right)
\label{3lgamma3g} {},       \\
\lb\gamma_{3}^{(2g)}\rb^{(4)}&=& 
        -\ca^4 \lb \f{252385}{1944} - \f{1045}{12} \zeta_{3}+ \f{111}{16} \zeta_{4}+ \f{5125}{48} \zeta_{5} \rb 
       + \dAAfNg \left(          \f{131}{36}    - \f{307}{6} \zeta_{3}     \right. \nonumber \\  & &\left.  
       - \f{335}{2} \zeta_{5}  \right) 
       + \sum\limits_i \Nfi \left\{ \tri   \left[
          - 46 \cfi^3                     
          + \ca \cfi^2 \lb \f{10847}{54}  + \f{980}{9} \zeta_{3}      - 240 \zeta_{5}  \rb \right.\right. \nonumber \\  & &\left.\left.
          - \ca^2 \cfi \lb \f{363565}{1944}  - \f{2492}{9} \zeta_{3}    + 126 \zeta_{4}    - 120 \zeta_{5} \rb   
          + \ca^3 \lb \f{1404961}{3888}      \right.\right.\right. \nonumber \\  & &\left.\left.\left.
          - \f{1285}{4} \zeta_{3}            + \f{387}{4} \zeta_{4}   + 110 \zeta_{5} \rb 
          \right] + d^{\sss{(4)}}_{\sss{FA,i}}   \left( - \f{512}{9}  + \f{1376}{3} \zeta_{3}   + 120 \zeta_{5}  \right)\right\}   \nonumber \\  & &
       + \sum\limits_{i,j} \Nfi \Nfj \left\{
        \tri \trj   \left[
           \cfj^2 \lb \f{304}{27}  + \f{128}{9} \zeta_{3} \rb
          - \cfi \cfj \lb \f{184}{3}  - 64 \zeta_{3} \rb                     \right.\right. \nonumber \\  & &\left.\left.
          - \ca \cfj \lb \f{15082}{243}  + \f{1168}{9} \zeta_{3}  - 48 \zeta_{4} \rb
          - \ca^2 \lb \f{41273}{486}  - \f{340}{9} \zeta_{3} + 36 \zeta_{4} \rb
          \right]                                \right. \nonumber \\  & &\left.
       +\dFFfijNg   \left(           \f{704}{9}          - \f{512}{3} \zeta_{3}          \right)\right\}        \nonumber \\  & &
       - \sum\limits_{i,j,k} \Nfi \Nfj \Nfk \tri \trj  \trk   \left[
           \f{1232}{243} \cfi
          + \ca \lb \f{1420}{243} - \f{64}{9} \zeta_{3} \rb
          \right]
\label{4lgamma3g} {}.
\eea 
From the ghost field strength renormalization constant $Z_{3}^{(2c)}$ we compute
\bea
\lb\gamma_{3}^{(2c)}\rb^{(1)}&=& 
       - \f{1}{2} \ca \label{1lgamma3c} {},       \\
\lb\gamma_{3}^{(2c)}\rb^{(2)}&=&    - \f{49}{24} \ca^2    
  +\f{5}{6} \ca \sum\limits_i \Nfi \tri   
         \label{2lgamma3c} {},       \\
\lb\gamma_{3}^{(2c)}\rb^{(3)}&=&  - \ca^3 \lb   \f{229}{27}   - \f{3}{4} \zeta_{3} \rb
       + \ca\sum\limits_i \Nfi \tri   \left[
             \cfi \lb \f{45}{4} - 12 \zeta_{3} \rb                 \right. \nonumber \\  & &\left.
          +  \ca \lb \f{5}{216} + 9 \zeta_{3} \rb
          \right]                                                
       +\f{35}{27} \ca \sum\limits_{i,j} \Nfi \Nfj \tri \trj 
\label{3lgamma3c} {},       \\
\lb\gamma_{3}^{(2c)}\rb^{(4)}&=& -\ca^4 \lb
       \f{256337}{3888} + \f{2485}{72} \zeta_{3} - \f{123}{32} \zeta_{4} - \f{4505}{96} \zeta_{5}      \rb
       + \dAAfNg   \left( \f{21}{8}- \f{299}{4} \zeta_{3}     \right. \nonumber \\  & &\left.
       + \f{265}{4} \zeta_{5}    \right)      
       + \sum\limits_i\Nfi \left\{ \tri \ca  \left[
          -  \cfi^2 \lb \f{271}{12}  + 74 \zeta_{3}  - 120 \zeta_{5}  \rb  \right.\right. \nonumber \\  & &\left.\left.
          + \ca \cfi \lb \f{22517}{432}  - 86 \zeta_{3}              + 69 \zeta_{4} - 60 \zeta_{5} \rb
          + \ca^2 \lb \f{449239}{7776}  + \f{2983}{24} \zeta_{3} \right.\right.\right. \nonumber \\  & &\left.\left.\left.
          - \f{423}{8} \zeta_{4}  - 55 \zeta_{5} \rb
          \right]   
          + \dFAfiNg   \left(48 \zeta_{3}          - 60 \zeta_{5}       \right) \right\} \nonumber \\  & &
       -\ca \sum\limits_{i,j}\Nfi \Nfj \tri \trj   \left[
           \cfj \lb \f{115}{27}  - 40 \zeta_{3}  + 24 \zeta_{4}  \rb \right. \nonumber \\  & &\left.
          + \ca \lb \f{8315}{972} + \f{86}{3} \zeta_{3}  - 18 \zeta_{4}  \rb    
          \right] \nonumber \\  & &
       + \sum\limits_{i,j,k}\Nfi \Nfj \Nfk \tri \trj \trk \ca  \left(           \f{166}{81}           - \f{32}{9} \zeta_{3}           \right)
\label{4lgamma3c} {}.
\eea
From the fermion field strength renormalization constant $Z_{2}^{(q,r)}$ we find 
\bea
\lb\gamma_{2}^{(q,r)}\rb^{(1)}&=& 
       \cfr  \label{1lgamma2qr} {},       \\
\lb\gamma_{2}^{(q,r)}\rb^{(2)}&=&        
       - \f{3}{2} \cfr^2  + \f{17}{2} \ca \cfr
       - 2\, \cfr \sum\limits_i\Nfi \tri  
         \label{2lgamma2qr} {},       \\
\lb\gamma_{2}^{(q,r)}\rb^{(3)}&=&  
        \f{3}{2} \cfr^3
          +\ca \cfr^2\lb - \f{143}{4} + 12 \zeta_{3} \rb
          + \ca^2 \cfr \lb \f{10559}{144} - \f{15}{2} \zeta_{3}\rb         \nonumber \\  & &
       - \cfr \sum\limits_i \Nfi \tri   \left(           6 \cfi          - 9 \cfr          + \f{1301}{36} \ca 
          \right) \nonumber \\  & &
       +  \f{20}{9} \cfr \sum\limits_{i,j} \Nfi \Nfj \tri \trj  
\label{3lgamma2qr} {},       \\
\lb\gamma_{2}^{(q,r)}\rb^{(4)}&=& 
       -\cfr^4\lb \f{1027}{8}           + 400 \zeta_{3}          - 640 \zeta_{5} \rb
          + \ca \cfr^3 \lb \f{5131}{12}          + 848 \zeta_{3}           - 1440 \zeta_{5} \rb            \nonumber \\  & &
          - \ca^2 \cfr^2 \lb \f{23777}{36}   + 214 \zeta_{3}    + 66 \zeta_{4} - 790 \zeta_{5} \rb
          + \ca^3 \cfr \lb \f{10059589}{15552}   \right. \nonumber \\  & &\left.
          - \f{1489}{24} \zeta_{3}      + \f{173}{4} \zeta_{4}   - \f{1865}{12} \zeta_{5} \rb
     - \dFAfNR   \left(   66  - 190 \zeta_{3} + 170 \zeta_{5}   \right)    \nonumber \\  & &
       +\sum\limits_{i} \Nfi \left\{ \tri \cfr   \left[
           3  \cfi^2
          + \cfr \cfi \lb 62   - 48 \zeta_{3} \rb
          - \cfr^2  \lb \f{119}{3}     + 16 \zeta_{3} \rb                      \right.\right. \nonumber \\  & &\left.\left.
          - \ca \cfi \lb \f{2945}{12} - 156 \zeta_{3} - 12 \zeta_{4} \rb
          + \ca \cfr \lb \f{1607}{9}   - 112 \zeta_{3}    + 24 \zeta_{4}  \right.\right.\right. \nonumber \\  & &\left.\left.\left.
          + 160 \zeta_{5} \rb  
          - \ca^2 \lb \f{1365691}{3888}  + \f{119}{3} \zeta_{3}  + 25 \zeta_{4} + 80 \zeta_{5} \rb
          \right]                
          + 128\, \dFFfiNR \right\} \nonumber \\  & &
       - \sum\limits_{i,j}\Nfi \Nfj \tri \trj \cfr   \left[           \f{92}{9} \cfr
          - \cfj \lb 44           - 32 \zeta_{3} \rb         \right. \nonumber \\  & &\left.
          - \ca  \lb \f{6835}{243}   + \f{112}{3} \zeta_{3} \rb
          \right] 
       + \f{280}{81} \cfr \sum\limits_{i,j,k}\Nfi \Nfj \Nfk \tri \trj \trk  
\label{4lgamma2qr} {}
\eea
for the anomalous dimension of a representation $r$ fermion field.

The fermion field-gauge boson-vertex renormalization constant $Z_{1}^{(q,r)}$ yields
\bea
\lb\gamma_{1}^{(q,r)}\rb^{(1)}&=&  
        \cfr  
       + \ca  \label{1lgamma1qr} {},       \\
\lb\gamma_{1}^{(q,r)}\rb^{(2)}&=&   
       - \f{3}{2}\cfr^2  
       + \f{17}{2} \ca \cfr  
       + \f{67}{24} \ca^2       
       - \sum\limits_i\Nfi \tri   \left(
           2 \cfr + \f{5}{6}\ca 
          \right) \label{2lgamma1qr} {},       \\
\lb\gamma_{1}^{(q,r)}\rb^{(3)}&=& 
        \f{3}{2} \cfr^3          
          - \ca \cfr^2 \lb \f{143}{4}          - 12 \zeta_{3} \rb
          + \ca^2 \cfr \lb \f{10559}{144}          - \f{15}{2} \zeta_{3} \rb  \nonumber \\  & &
          + \ca^3\lb \f{10703}{864}     + \f{3}{4} \zeta_{3} \rb                
        + \sum\limits_{i}\Nfi \tri   \left[
          - 6 \cfr \cfi
          + 9 \cfr^2                                        \right.     \nonumber \\  & & \left.
          - \ca \cfi \lb \f{45}{4}  - 12 \zeta_{3} \rb    
          - \f{1301}{36} \ca \cfr                             
          - \ca^2 \lb \f{205}{108}   + 9 \zeta_{3} \rb
          \right]                                         \nonumber \\  & &     
       + \sum\limits_{i,j}\Nfi \Nfj \tri \trj   \left(
           \f{20}{9} \cfr
          - \f{35}{27} \ca
          \right)   
\label{3lgamma1qr} {},       \\
\lb\gamma_{1}^{(q,r)}\rb^{(4)}&=& 
       - \cfr^4 \lb \f{1027}{8}   + 400 \zeta_{3} - 640 \zeta_{5} \rb
          + \ca \cfr^3 \lb \f{5131}{12}   + 848 \zeta_{3}   - 1440 \zeta_{5} \rb       \nonumber \\  & & 
          - \ca^2 \cfr^2 \lb \f{23777}{36}   + 214 \zeta_{3} + 66 \zeta_{4} - 790 \zeta_{5} \rb
          +\ca^3 \cfr \lb \f{10059589}{15552}    \right.     \nonumber \\  & & \left.
          - \f{1489}{24} \zeta_{3}  + \f{173}{4} \zeta_{4}  - \f{1865}{12} \zeta_{5} \rb 
          + \ca^4\lb \f{350227}{3888}  + \f{2959}{72} \zeta_{3}   - \f{111}{32} \zeta_{4}     - \f{5125}{96} \zeta_{5} \rb \nonumber \\  & & 
       - \dAAfNg   \left( \f{21}{8}   - \f{367}{4} \zeta_{3}  + \f{335}{4} \zeta_{5} \right)
       - \dFAfNR   \left( 66    - 190 \zeta_{3}   + 170 \zeta_{5}    \right) \nonumber \\  & &        
       + \sum\limits_{i} \Nfi \left\{ \tri   \left[
          3 \cfr \cfi^2
          + \cfr^2 \cfi\lb 62    - 48 \zeta_{3} \rb
          - \cfr^3\lb \f{119}{3}    + 16 \zeta_{3} \rb    \right.\right.    \nonumber \\  & & \left.\left.
          + \ca \cfi^2\lb \f{271}{12} \   + 74 \zeta_{3} - 120 \zeta_{5} \rb
          - \ca \cfr \cfi \lb \f{2945}{12}    - 156 \zeta_{3} - 12 \zeta_{4} \rb \right.\right.    \nonumber \\  & & \left.\left.
          + \ca \cfr^2 \lb \f{1607}{9}    - 112 \zeta_{3}    + 24 \zeta_{4}   + 160 \zeta_{5} \rb
          - \ca^2 \cfi \lb \f{34109}{432}   - 102 \zeta_{3} + 63 \zeta_{4} \right.\right.\right.   \nonumber \\  & & \left.\left.\left.
          - 60 \zeta_{5} \rb 
          - \ca^2 \cfr\lb \f{1365691}{3888}   + \f{119}{3} \zeta_{3} + 25 \zeta_{4} + 80 \zeta_{5} \rb
          - \ca^3 \lb \f{473903}{7776}   + \f{3311}{24} \zeta_{3} \right.\right.\right.   \nonumber \\  & & \left.\left.\left.
          - \f{387}{8} \zeta_{4} - 55 \zeta_{5} \rb
          \right]       
          + 128\,\dFFfiNR 
          - \dFAfiNg   \left(48 \zeta_{3}  - 60 \zeta_{5}    \right)
          \right\}   \nonumber \\  & &
       + \sum\limits_{i,j} \Nfi \Nfj \tri \trj   \left[
          \cfr \cfj    \lb 44   - 32 \zeta_{3} \rb
          - \f{92}{9} \cfr^2
          + \ca \cfj \lb \f{115}{27}    - 40 \zeta_{3}  \right.\right.   \nonumber \\  & & \left.\left.
          + 24 \zeta_{4} \rb      
          + \ca \cfr \lb \f{6835}{243}    + \f{112}{3} \zeta_{3} \rb
          + \ca^2 \lb \f{6307}{972}  + \f{94}{3} \zeta_{3} - 18 \zeta_{4} \rb
          \right]  \nonumber \\  & &
       + \sum\limits_{i,j,k}\Nfi \Nfj \Nfk \tri \trj  \trk \left[
          \f{280}{81} \cfr
          - \ca \lb \f{166}{81}  - \f{32}{9} \zeta_{3} \rb          
          \right]
\label{4lgamma1qr} {}
\eea
for each representation $r$ and the ghost-gauge boson-vertex renormalization constant $Z_{1}^{(ccg)}$ yields
\bea
\lb\gamma_{1}^{(ccg)}\rb^{(1)}&=& 
       \f{1}{2}\ca  \label{1lgamma1ccg} {},       \\
\lb\gamma_{1}^{(ccg)}\rb^{(2)}&=&        \f{3}{4} \ca^2 \label{2lgamma1ccg} {},       \\
\lb\gamma_{1}^{(ccg)}\rb^{(3)}&=& \f{125}{32} \ca^3      
   - \f{15}{8} \ca^2 \sum\limits_i \Nfi \tri   
\label{3lgamma1ccg} {},       \\
\lb\gamma_{1}^{(ccg)}\rb^{(4)}&=& 
       \ca^4 \lb \f{46945}{1944} + \f{79}{12} \zeta_{3}  + \f{3}{8} \zeta_{4}    - \f{155}{24} \zeta_{5} \rb  
       + \dAAfNg   \left(   17 \zeta_{3}     - \f{35}{2} \zeta_{5}         \right) \nonumber \\  & &
       - \sum\limits_{i}\Nfi \tri  \ca^2  \left[
          \cfi \lb \f{161}{6} - 16 \zeta_{3} - 6 \zeta_{4} \rb
          + \ca \lb \f{3083}{972}  \right.\right.\nonumber \\  & & \left. \left.
          + \f{41}{3} \zeta_{3} + \f{9}{2} \zeta_{4} \rb   
          \right]
       - \sum\limits_{i,j}\Nfi \Nfj \tri \trj \ca^2  \left(  \f{502}{243}  - \f{8}{3} \zeta_{3}      \right)
\label{4lgamma1ccg} {}.
\eea
Finally, the mass anomalous dimension computed from $Z_{m}^{(q,r)}$ is found to be
\bea
\lb\gamma_{m}^{(q,r)}\rb^{(1)}&=& 
       3\, \cfr    \label{1lgammamqr} {},       \\
\lb\gamma_{m}^{(q,r)}\rb^{(2)}&=&        \f{3}{2} \cfr^2 + \f{97}{6}\ca \cfr 
        - \f{10}{3} \cfr \sum\limits_i\Nfi \tri  \label{2lgammamqr} {},       \\
\lb\gamma_{m}^{(q,r)}\rb^{(3)}&=&      
        \f{129}{2} \cfr^3
          - \f{129}{4} \ca \cfr^2
          + \f{11413}{108} \ca^2 \cfr      \nonumber \\  & &                     
       - \cfr \sum\limits_{i} \Nfi \tri  \left[           \cfr 
          + \cfi \lb 45 - 48 \zeta_{3} \rb
          + \ca \lb \f{556}{27}   + 48 \zeta_{3} \rb
          \right]                                               \nonumber \\  & & 
       - \f{140}{27} \cfr \sum\limits_{i,j} \Nfi \Nfj \tri \trj  
\label{3lgammamqr} {},       \\
\lb\gamma_{m}^{(q,r)}\rb^{(4)}&=& 
          - \cfr^4\lb \f{1261}{8}    + 336 \zeta_{3} \rb
          + \ca \cfr^3\lb \f{15349}{12}    + 316 \zeta_{3} \rb 
          - \ca^2 \cfr^2 \lb \f{34045}{36}  \right. \nonumber \\  & & \left. 
          + 152 \zeta_{3}    - 440 \zeta_{5} \rb
          + \ca^3 \cfr \lb \f{70055}{72}  + \f{1418}{9} \zeta_{3}  - 440 \zeta_{5} \rb 
      - \dFAfNR   \left(           32          - 240 \zeta_{3}          \right) \nonumber \\  & &
       + \sum\limits_{i}\Nfi \left\{ \tri \cfr   \left[
          \cfi^2 \lb \f{271}{3}   + 296 \zeta_{3}      - 480 \zeta_{5} \rb
          - \cfr \cfi \lb 38    - 48 \zeta_{3} \rb                \right.\right. \nonumber \\  & & \left.\left.
          - \cfr^2 \lb \f{437}{3}    - 208 \zeta_{3} \rb
          - \ca  \cfi \lb \f{13106}{27}      - 592 \zeta_{3} + 264 \zeta_{4} - 240 \zeta_{5} \rb  \right.\right. \nonumber \\  & & \left.\left.
          + \ca \cfr \lb \f{1429}{9}   - 224 \zeta_{3}   - 160 \zeta_{5} \rb
          - \ca^2  \lb \f{65459}{162}  + \f{2684}{3} \zeta_{3}    - 264 \zeta_{4}   - 400 \zeta_{5} \rb
          \right]    \right. \nonumber \\  & & \left. 
          + \dFFfiNR   \left(    64          - 480 \zeta_{3}          \right) \right\} 
       + \cfr \sum\limits_{i,j}\Nfi \Nfj \tri \trj  \left[
            \cfj \lb \f{460}{27}     - 160 \zeta_{3}    + 96 \zeta_{4} \rb \right. \nonumber \\  & & \left.
          - \f{52}{9} \cfr                     
          + \ca \lb \f{1342}{81}   + 160 \zeta_{3}  - 96 \zeta_{4} \rb
          \right]                           \nonumber \\  & &
       - \sum\limits_{i,j,k}\Nfi \Nfj \Nfk\tri \trj  \trk  \cfr  \left(  \f{664}{81} - \f{128}{9} \zeta_{3}  \right)
  \label{4lgammamqr} {}.
\eea

We checked that the well known relations
\bea \f{\beta(a)}{a} &=& 2 \gamma_1^{(ccg)}(a,\lambda) - 2 \gamma_3^{(2c)}(a,\lambda)  - \gamma_3^{(2g)}(a,\lambda) {}, \\
\f{\beta(a)}{a} &=& 2 \gamma_1^{(q,r)}(a,\lambda) - 2 \gamma_2^{(q,r)}(a,\lambda)  - \gamma_3^{(2g)}(a,\lambda) {}
\eea
are fulfilled with the $\beta$-function from \cite{Zoller:2016sgq}. This is also true if we include the full dependence on the gauge parameter 
$\xi=1-\lambda$ in the anomalous dimensions. This dependence cancels in the $\beta$-function. We provide renormalization constants and 
anomalous dimensions with the full gauge dependence in the attached files, which can be downloaded with the source files of this paper from 
\texttt{www.arxiv.org}. We compared these fully $\xi$-dependent results with \cite{Luthe:2017ttc} for one fermion representation and find full agreement.
\section{Conclusions \label{last}}

We have presented analytical results for the field anomalous dimensions $\gamma_{3}^{(2g)}$, $\gamma_{3}^{(2c)}$, $\gamma_{2}^{(q,r)}$,
the vertex anomalous dimensions $\gamma_{1}^{(ccg)}$ and $\gamma_{1}^{(q,r)}$ and the mass anomalous dimension $\gamma_{m}^{(q,r)}$ 
in a QCD-like model with arbitrarily many fermion representations and with the full dependence on the 
gauge parameter $\xi$.

\section*{Acknowledgements}
The work by K.~G.~Chetykin  was supported by the Deutsche Forschungsgemeinschaft through CH1479/1-1 and in part by the German
Federal Ministry for Education and Research BMBF through Grant No.~05H2015.
The work by M.~F.~Zoller was supported by the Swiss National Science Foundation (SNF) under contract BSCGI0\_157722.

\bibliographystyle{JHEP}

\bibliography{LiteraturSM_2016_v4}

\end{document}